\documentclass[prd, 11pt, a4paper, showpacs, notitlepage, dvipsnames, reprint, twocolumn, preprintnumbers, nofootinbib, floatfix]{revtex4-1}

\usepackage[toc,page]{appendix}

\usepackage{geometry}
\usepackage[utf8x]{inputenc}
\usepackage{makecell}
\usepackage{float}
\usepackage{tabularx}
\usepackage[activate={true,nocompatibility},final,tracking=true,kerning=true,factor=1500,stretch=10,shrink=10]{microtype}
\usepackage{graphicx}
\usepackage{bm, amssymb, amsmath, amsfonts}
\usepackage{multirow}
\usepackage{xcolor}

\usepackage{url}

\usepackage{url}
\usepackage[breaklinks]{hyperref}
\usepackage{todonotes}

\usepackage{hyperref}
\hypersetup{
colorlinks = true, 	    
linkcolor = magenta,	
citecolor = blue,		
}

\usepackage[linesnumbered,ruled,lined,boxed]{algorithm2e}
\usepackage[noend]{algpseudocode}

\SetAlCapSty{xAlCapSty}

\SetCommentSty{mycommfont}

\SetNlSty{mynlfont}{}{} 

\RestyleAlgo{algoruled}

\setcitestyle{numbers,square}

\begin{document}

\title{A convolutional neural network to distinguish glitches from minute-long gravitational wave transients}
\author{Vincent Boudart$^{1}$} \email[]{vboudart@uliege.be}

\affiliation{${}^1$ STAR Institute, Bâtiment B5, Université de Liège, Sart Tilman B4000 Liège, Belgium}

\begin{abstract}
\noindent
Gravitational wave bursts are transient signals distinct from compact binary mergers that arise from a wide variety of astrophysical phenomena. Because most of these phenomena are poorly modeled, the use of traditional search methods such as matched filtering is excluded. Bursts include short ($<$10 seconds) and long (from 10 to a few hundreds of seconds) duration signals for which the detection is constrained by environmental and instrumental transient noises called glitches. Glitches contaminate burst searches, reducing the amount of useful data and limiting the sensitivity of current algorithms. It is therefore of primordial importance to locate and distinguish them from potential burst signals. In this paper, we propose to train a convolutional neural network to detect glitches in the time-frequency space of the cross-correlated LIGO noise. We show that our network is retrieving more than 95$\%$ of the glitches while being trained only on a subset of the existing glitch classes highlighting the sensitivity of the network to completely new glitch classes.
\end{abstract}
\maketitle 


\section*{INTRODUCTION}

Gravitational waves (GW) have been detected on September 14, 2015 \cite{first_detection} by the Advanced LIGO \cite{aLIGO} detectors, revealing the collision of two black holes for the first time. Since then, the Advanced LIGO and the Advanced Virgo \cite{Virgo} detectors have observed more than 90 compact binary coalescence (CBC) events \cite{GWTC3}, among which black hole-neutron star \cite{NSBH} and binary neutron star collisions \cite{neutron_star}. In light of the planned sensitivity improvement of the Advanced LIGO and Advanced Virgo detectors, a new family of gravitational wave sources, known as unmodeled GW transients or bursts, is a prime target candidate for the next observing run. Bursts include a wide range of astrophysical phenomena for which accurate waveforms are not accessible. The computational resources required to build a template bank covering a wide range of complex and highly turbulent events prevents us to use matched filtering methods as in CBC searches \cite{matched_filtering}. Some of the expected progenitors of gravitational wave transients are supernovae \cite{supernovae}, fallback accretion events \cite{fallback_accretion_NS}, accretion-disk instabilities \cite{adi}, nonaxisymmetric deformations in magnetars \cite{magnetars}, accretion-disk instabilities \cite{adi} as well as gamma-ray bursts \cite{GRB}. Two classes of bursts are identified: short ($<10$ seconds) and long (from 10 to a few hundreds of seconds). In this paper, we present a new machine learning tool that complements our previous work \cite{mypaper} and discriminates transient noises happening in the detectors from long-duration burst signals. \\

The main approach to detect burst events while making minimal assumptions on the targeted signals relies on the excess-of-power method. It consists in searching for excess of power in the time-frequency space of single or multiple detector data, i.e. to find narrow time-evolving frequency curves. This problem has already been tackled by different groups who built the current generation of pipelines, namely PySTAMPAS \cite{PySTAMPAS}, cocoA \cite{cocoA}, the two different versions of STAMP-AS, Zebragard and Lonetrack \cite{STAMP_1,STAMP_2}, the long-duration configuration of coherent WaveBurst (cWB) \cite{cWB} and X-SphRad \cite{Fays}. \\

One of the main hindrance in burst searches is glitches. Glitches are transient noises caused by instrumental or environmental sources \cite{noise_characterization2, noise_characterization} that appear in the detector data in large quantities. Several families of glitches have been reported \cite{GravitySpy}, showing different time-frequency morphologies. Glitches limit the sensitivity of the searches and can hinder GW detections. Therefore, all of the aforementioned pipelines deal with glitches either in pre- or post-processing steps. In a previous work \cite{mypaper}, we trained a neural network with chirp signals having random parameters and showed that our methodology can be used to detect minute-long GW transients. However, it can also recover glitches fairly and a visual inspection is needed to discriminate them from chirp signals. This work aims at removing the false-alarms caused by glitches through a convolutional neural network. \\

Convolutional neural networks (CNNs) have been recently used in Burst detection \cite{short_duration_burst}. The authors in \cite{short_duration_burst} have built a 1 dimensional CNN to detect generic short duration signals from the strain data of LIGO and Virgo detectors. CNNs have shown promising results in the identification and classification of GW bursts from supernovae \cite{Melissa, CCSN_CNN}, in the detection of binary black hole mergers \cite{CBC_cnn} as well as long-duration transients from isolated neutron stars \cite{continuous_CNN} or as early alert systems for binary neutron star collisions \cite{Baltus}. CNNs are widely used for pattern recognition \cite{mypaper, YOLO} and classification tasks \cite{GoogleNet, AlexNet, EfficientNet}. Their powerful capability to identify shapes and structures has lead to the definition of Generative Adversarial Networks \cite{Original_GAN}, allowing to generate new samples by learning the underlying distribution of the original data. \\

In Sec. \ref{section:methodology}, we describe how glitches have been selected to constitute the training set and how we highlight them in the cross-correlated TF maps. Details about the architecture of our classifier and the training method are given in Sec. \ref{section:neural_network}. We then show the results of the training in Sec. \ref{section:results}. Section \ref{section:burst_search} is dedicated to large scale tests comparable to the analyses conducted during burst searches. Future prospects and conclusions are given in Sec. \ref{section:conclusion}.

\section{Methodology}\label{section:methodology}

Our search for minute-long bursts is based on the excess-of-power method \cite{basics_GW}. We make use of correlated spectrograms, also referred to as time-frequency (TF) maps, as described in \cite{mypaper}. In order to distinguish glitches from possible burst signals, we will train a neural network to identify them in the spectrograms. As both can be present in a single TF map, we need to consider the following cases : (1) a glitch is present in the map, (2) a burst signal is present in the map, (3) both or (4) none of them show up in the spectrogram. Accordingly we will build 4 different data sets to include all the possible scenarios in the training phase. \\

The fourth scenario consists in building a dataset with background TF maps. The data from Hanford (H1) and Livingston (L1) from the first half of the third observing run (O3a) are first whitened \cite{whitening} prior to be correlated. Using time-slides \cite{time_slides}, we then generate 10000 spectrograms with a time resolution of 6 seconds and a frequency resolution of 2 Hz. As the TF maps span 1000 seconds and 2048 Hz, their size is 166$\times$1025. Since we aim to apply our classifier on ALBUS' output, the size of the TF maps is chosen to be identical to \cite{mypaper}.

\subsection{Chirp generation}\label{subsection:chirp}
\begin{figure*}[htb]
    \centering
    \includegraphics[trim={0cm 0cm 0cm 0cm}, clip, scale=0.56]{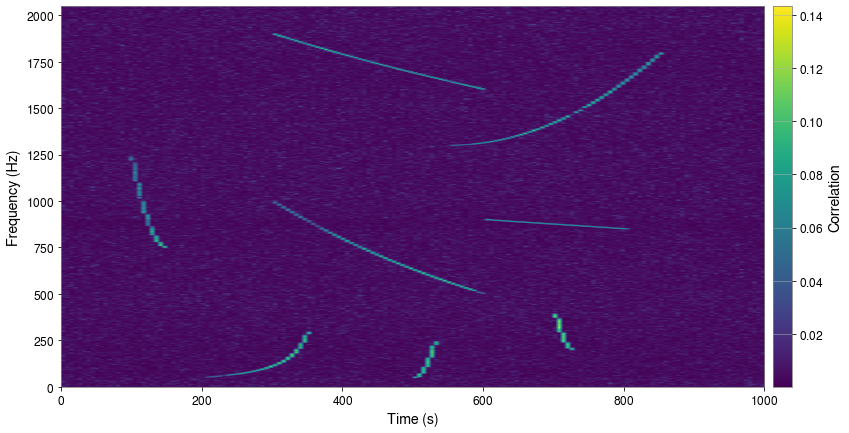}
    \caption{Examples of chirp signals.}
    \label{example_chirps}
\end{figure*}

A methodology to recognize minute-long burst signals using machine learning techniques with very few assumptions has been proposed in our previous work \cite{mypaper}. This approach consists in using the Scipy library \cite{scipy} to generate chirp signals in the time domain with random parameters, covering the whole time-frequency parameter space. Figure \ref{example_chirps} shows some examples of generated chirps. As has been shown \cite{mypaper}, this allows to train a neural network with no prior assumption on the targeted signals while confidently identifying minute-long burst models. Chirps are injected into noise with 9 levels of visibility, defined as :

\begin{equation}
    V = \sum_{i,j}\, \big( S_{ij}\,-\,N_{ij} \big)
    \label{visibility}
\end{equation}

\noindent
where $N_{ij}$ is a noise-only spectrogram and $S_{ij}$ refers to the same spectrogram in which a signal has been injected. The sum is carried over all the pixels $(i,j)$ in the map. The definition of the visibility is particularly useful to ensure chirps to be visible in the TF maps, preventing the network to be fooled during the training phase. The visibility can also be seen as a measure of the anomalousness of the input TF maps. We choose 9 intensity levels in order to cover a quite large intensity range, as seen in Figure \ref{visibility_levels} in the Appendix. We use this intensity criterion to build our second dataset, containing 10000 samples.

\subsection{Glitch selection}
During the second observing run (O2), glitches happened roughly at a rate of 1 every min in the detectors \cite{detchar2021}. Although it amounts to a considerable volume of contaminated data, they barely show up in cross-correlated spectrograms. Indeed, both glitches have to fall into overlapping time bins while showing a sufficiently high signal-to-noise ratio (SNR) and sharing some frequency bandwidth. Even if these conditions greatly reduce the amount of glitches that contaminate our search, several thousands of glitches can be found out of a couple of millions of TF maps generated during the background searches. \\

To constitute our dataset with glitches, we need a way to inject several glitch classes into time frequency maps. However, the only tool that is currently available to produce realistic glitches can only generate \textit{blip} glitches \cite{gengli_first}. Blips are one the 23 classes that have been characterized by Gravity Spy \cite{GravitySpy, 23classes}. They have a frequency between 0 and 256 Hz \cite{blip_ref_1, blip_ref_2} which would limit the detection bandwidth of the classifier if used exclusively in our dataset. Therefore, we have to rely on the glitches detected so far to constitute the training set. We thus select glitches that have been recorded by Gravity Spy during O3a \cite{data_release_GS}. We load the data around the GPS time of the chosen glitch in each single detector (Hanford H1 and Livingston L1) and shift them so that they fall into the same time bin. In this way, we maximize the probability of finding cross-correlated glitches that appear clearly in the TF maps. Moreover, glitches showing higher SNR do not always lead to stronger cross-correlated signals in the TF maps. To circumvent these problems, we choose 7 glitch classes with SNR ranging from 20 to 10000 in both Hanford and Livingston data. This will ensure some variability in the results of the cross-correlation. Table \ref{table1} summarizes the useful information. \\


\begin{table}[!htb]
   \centering
   \renewcommand{\arraystretch}{1.5}
   \begin{tabular}{|c|c|}
     \hline
       \textbf{Glitch classes} & \begin{tabular}{@{}c@{}} Blip, Low Frequency Burst, \\ Scattered Light, Tomte, \\ Whistle, Extremely Loud, \\ Koi Fish \end{tabular} \\
     \hline
       \textbf{SNR ranges} & \begin{tabular}{@{}c@{}} 20-30, 30-40, 40-50, 50-100, \\ 100-150, 150-200, 200-300, \\ 300-500, 500-10000 \end{tabular} \\
     \hline
       \textbf{Number per range} & 30 (if possible) \\
     \hline
       \textbf{Injection time} & between 50s and 950s \\
     \hline
       \textbf{Total} & \textbf{H1:} 1110 \ \textbf{L1:} 1260 \\
     \hline
   \end{tabular}
   \caption{Information about the glitches selected from H1 and L1.}
   \label{table1}
\end{table}

\begin{figure}[!htb]
    \centering
    \begin{tabular}{c}
        \includegraphics[scale=0.27, trim={0.1cm 0cm 0cm 0cm}, clip]{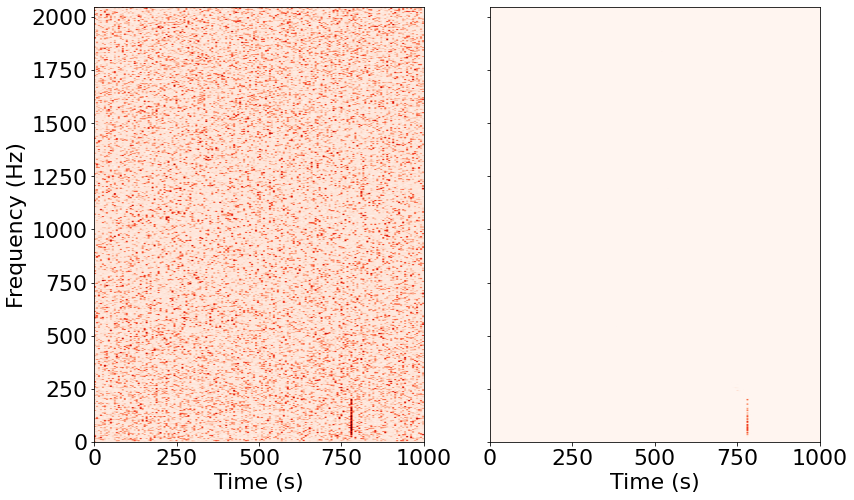}
    \end{tabular}
    \begin{tabular}{c}
        \includegraphics[scale=0.27, trim={0.1cm 0cm 0cm 0cm}, clip]{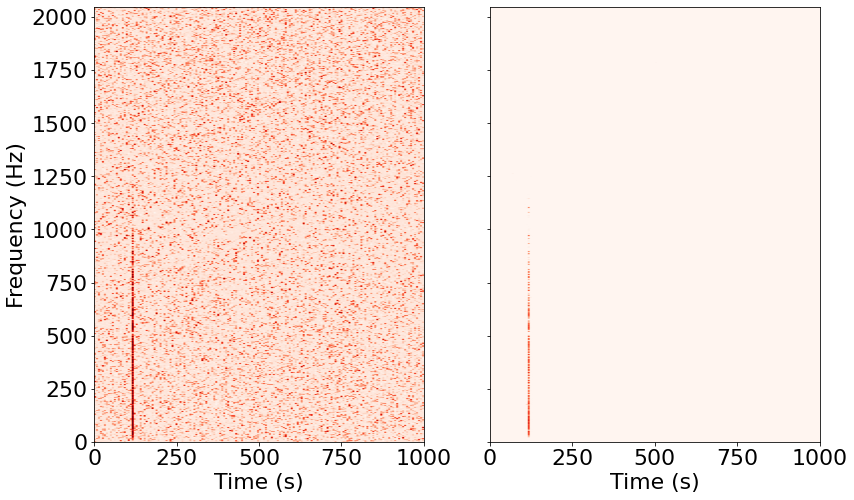}
    \end{tabular}
    \caption{Examples of correlated glitches with different anomaly scores. The left panel shows the generated spectrogram while the right panel shows the output of ALBUS. The top and bottom glitches have an anomaly score of 21 and 144 respectively.}
    \label{example_glitches}
\end{figure}

The total number of selected glitches is 1110 for H1 and 1260 for L1. Then, we randomly choose one glitch from each detector and build the resulting time-frequency map. We reproduce this procedure 50000 times. To evaluate if the cross-correlation of the chosen glitches has lead to a visible glitch in the output spectrogram, we employ ALBUS, the neural network dedicated to burst detection \cite{mypaper}. We showed that ALBUS can recover glitches as well as chirp signals. We use its output map to introduce a score quantifying the anomalousness present in the original spectrogram, called anomaly score (AS). This score is defined as :

\begin{equation}
    AS = \sum_{i,j}\, O_{i,j} \,\,\, \mathrm{if}\, O_{i,j}\,>\,0.5\,\max(O)
    \label{anomaly_score}
\end{equation}

where $O$ is the ALBUS output map and $i$ and $j$ indicate the time and frequency dimensions. The anomaly score can be thought of the sum over the pixels remaining after applying an intensity cut to the output map. This threshold has been chosen to exclude all the values close to zero, as they are quite numerous given the size of the TF maps and can have an impact on the final anomaly score. The anomaly score can also be used to rank detected signals as seen in Figure \ref{example_glitches} where an extended glitch shows a higher score compared to a glitch that spends a small frequency range.\\

\begin{figure}[!htb]
    \centering
    \includegraphics[trim={0.1cm 0cm 0cm 0cm}, clip, scale=0.35]{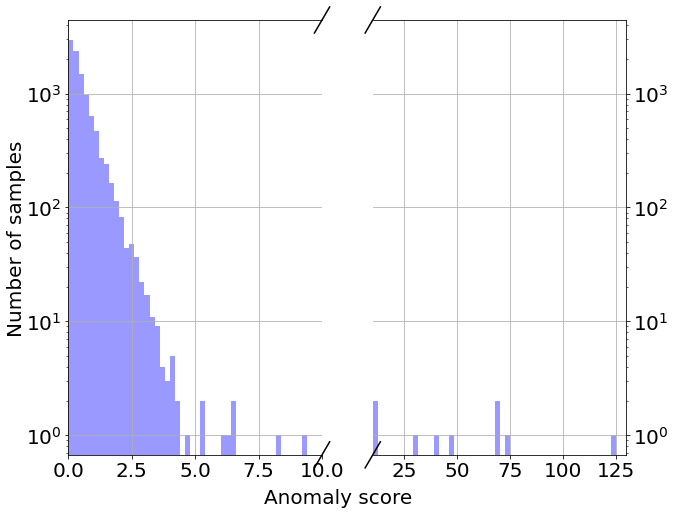}
    \caption{Histogram of the anomaly scores for 10000 background TF maps.}
    \label{as_background}
\end{figure}

After visual inspection, background maps without any glitch have a maximum score around 6.5, as seen in Figure \ref{as_background} where 10000 images have been processed. All the 15 background images with scores above 8 show a correlated glitch. We thus set the threshold to confirm the presence of a correlated glitch to 8 in order to leave a sufficient margin between high-noise level TF maps and those containing glitches. After applying this threshold to our 50000 spectrograms, we end up with only 4744 maps showing correlated glitches. The dataset has been drastically reduced but it is still sufficient to achieve a well-behaved training. 

\subsection{Combined dataset}
The procedure to generate spectrograms containing a chirp and a glitch is very similar to the method described in the previous subsection. We generate 45000 spectrograms with the glitches selected in Table \ref{table1}. Then, we inject chirp signals with 9 levels of visibility, as in subsection \ref{subsection:chirp}. Once the signal has been added to the map, we process the latter with ALBUS. \\

At this stage, we cannot rely on the anomaly score as it is defined. The chirp signals will also be recovered and contribute to the anomaly score of the map, which can hide the presence of a glitch. However, as we know where the chirp is injected, we can discard the corresponding pixels in the output map. This is done by masking the pixels corresponding to the footprint of the injected chirp. Then, the anomaly score is still relevant to assess whether or not a glitch is present in the maps. Out of the 45000 maps, 6068 actually pass the threshold and contain a correlated glitch and an injected chirp. This thorough check for glitches is important in light of the training approach explained in the next section. \\

\begin{figure*}[htb]
 \centering
 \includegraphics[trim={0cm 0cm 0cm 0cm}, clip, scale=0.6]{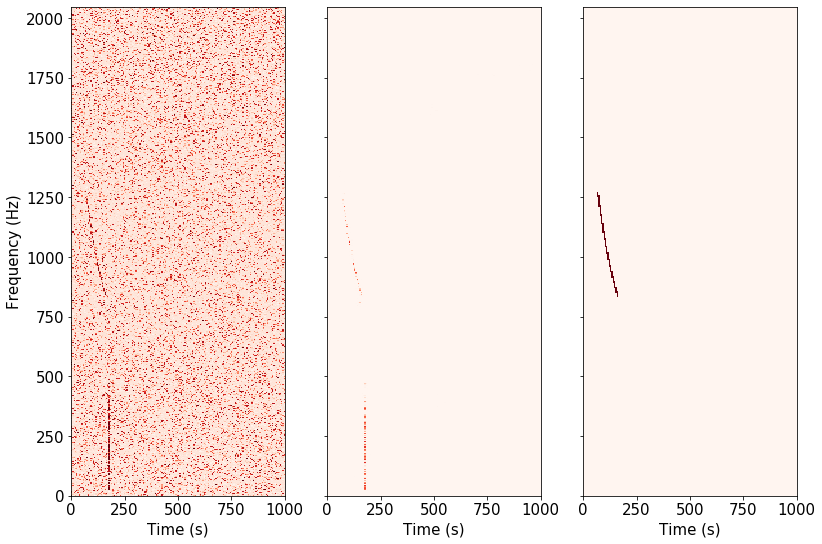}
 \caption{Example of TF map showing a glitch and a chirp signal. The left and center panels illustrate respectively the generated spectrogram and the output of ALBUS. The right panel corresponds to the chirp mask that is used to cancel out the contribution of the chirp in the estimated anomaly score.}
 \label{example_chirp_glitch}
\end{figure*}

Figure \ref{example_chirp_glitch} shows an example of spectrogram containing a chirp and a correlated glitch. The anomaly score of the output map is 49.6 while it drops to 34.5 when the chirp pixels are masked out.

\section{Machine Learning}\label{section:neural_network}

\begin{figure*}[!htb]
    \centering
    \includegraphics[trim={0cm 0cm 0cm 0cm}, clip, scale=0.58]{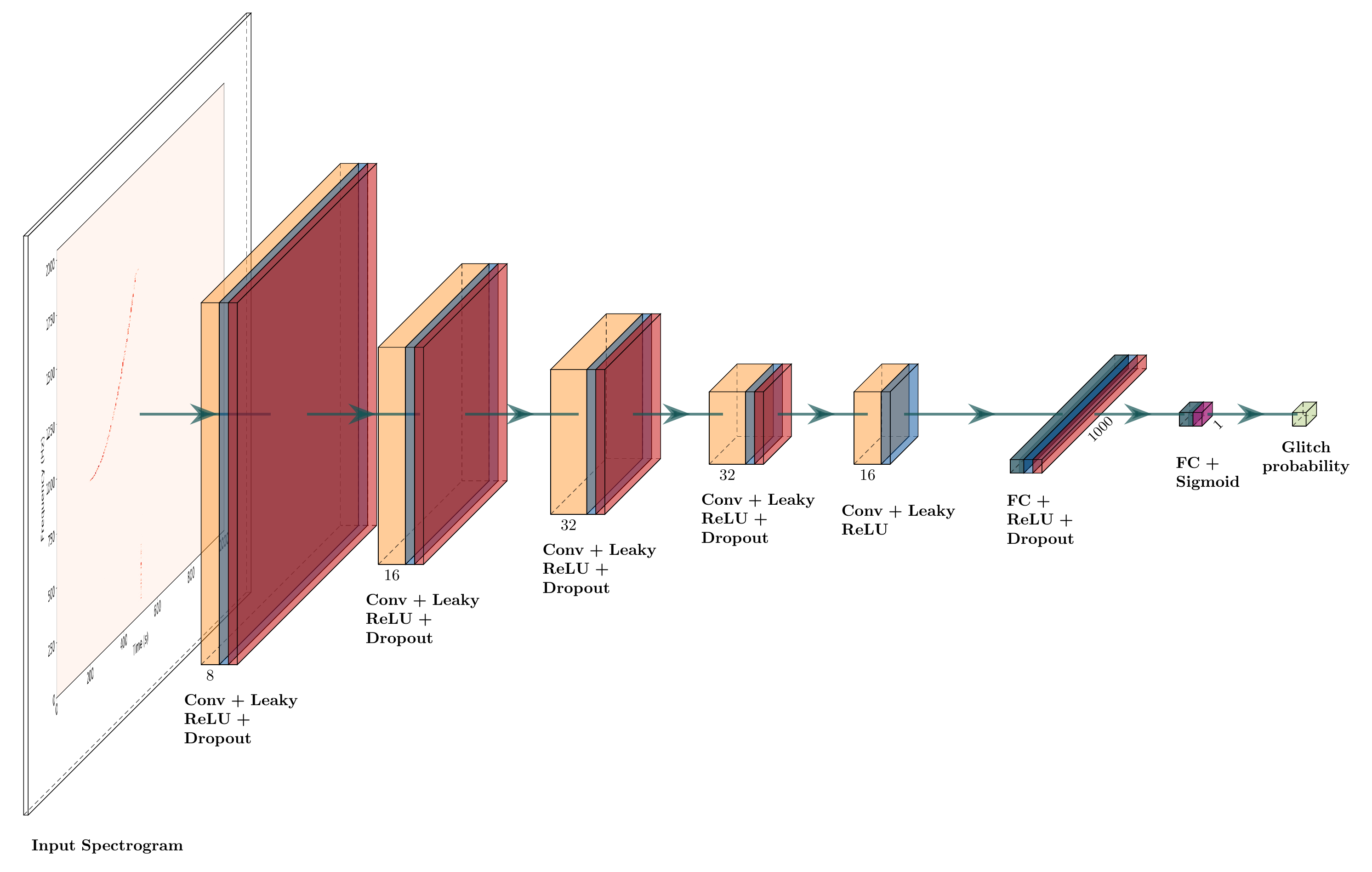}
    \caption{Architecture of the CNN. \textit{Conv.} and \textit{FC} stand respectively for convolution and fully-connected layers.}
    \label{architecture classifier}
\end{figure*}

In this work, we use a CNN to assess if a glitch is present in the time-frequency maps. For this, we feed the output of ALBUS \cite{mypaper} to a CNN, predicting a glitch probability. The full architecture can be seen in Figure \ref{architecture classifier}. The network is composed of two parts. The first part is fully convolutional and acts as a feature extractor. Then, a fully connected network uses these features to evaluate a glitch probability. The sigmoid activation function is used to obtain an output value between 0 and 1. The hyperparameters of the network have been chosen via trial and error. We add dropout \cite{dropout} to every convolution layer and the first dense layer with a probability of 30$\%$. Table \ref{hyperparameters} shows an exhaustive list of the parameters used across all layers. \\

\begin{table}[!htb]
   \centering
   \renewcommand{\arraystretch}{1.5}
   \begin{tabular}{|c|c|c|c|c|c|}
     \hline
        & \textbf{Nb of filters} & \textbf{Kernel size} & \textbf{Stride} & \textbf{Padding} \\
     \hline
       \textbf{Conv. 1} & 8 & 7$\times$7 & 2$\times$2 & 0$\times$0 \\
     \hline
       \textbf{Conv. 2} & 16 & 7$\times$7 & 1$\times$1 & 0$\times$0 \\
     \hline
       \textbf{Conv. 3} & 32 & 5$\times$5 & 2$\times$2 & 0$\times$0 \\
     \hline
       \textbf{Conv. 4} & 32 & 5$\times$5 & 1$\times$1 & 0$\times$0 \\
     \hline
       \textbf{Conv. 5} & 16 & 3$\times$3 & 2$\times$2 & 0$\times$0 \\
     \hline
       \textbf{FC 1} & 29280 & / & / & / \\
     \hline
       \textbf{FC 2} & 1000 & / & / & / \\
     \hline
   \end{tabular}
   \caption{Hyperparameters used in the architecture of our classifier.}
   \label{hyperparameters}
\end{table}

The training procedure is straightforward. Every single TF map is passed through the network with a glitch label, as summarized in Table \ref{label training}. The binary cross entropy (BCE) loss is applied between the predicted and real label:

\begin{equation}
    L = BCE\big( L_g, P_g \big)
\end{equation}

\noindent
with BCE being defined as :
\begin{equation}
    BCE(x,y) = y\,log(x) + (1-y)\,log(1-x)
\end{equation}

\noindent
where $L_g$ stands for the glitch label while $P_g$ is the predicted glitch probability.

\begin{table}[!htb]
   \centering
   \renewcommand{\arraystretch}{1.5}
   \begin{tabular}{|c|c|}
     \hline
     \textbf{TF maps} & \textbf{Glitch label} \\
     \hline
        Background & 0 \\
     \hline
        Chirp & 0 \\
     \hline
        Glitch & 1 \\
     \hline
        Combined & 1 \\
     \hline
   \end{tabular}
   \caption{Labels used for the training of the classifier.}
   \label{label training}
\end{table}

\section{Results}\label{section:results}
\subsection{Training}

\begin{figure}[htb]
    \centering
    \includegraphics[trim={0cm 0cm 0cm 0cm}, clip, scale=0.34]{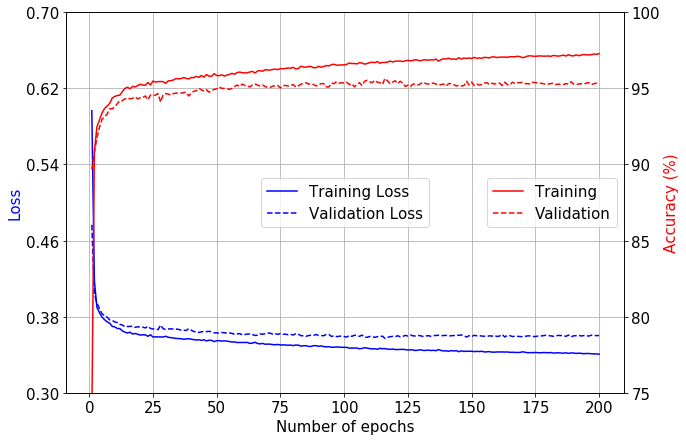}
    \caption{Loss and accuracy of the network for a training of 200 epochs.}
    \label{loss training}
\end{figure}

We select 4000 TF maps in each category, amounting to 16000 images for our dataset. A validation set of $20\%$ is used throughout the training. We use the Adamax optimizer, a variant of Adam \cite{Adam}, with a weight decay of $10^{-5}$ and a learning rate of $3\,10^{-5}$. The batch size is set to 32. The evolution of the loss and the accuracy to predict the glitch labels are shown in Figure \ref{loss training}. Both the training and validation losses behave smoothly during the training. We stop the training when the validation loss starts to rise again, indicating that the network starts to overfit the data. At the same time, the accuracy reaches a plateau and no further progress is observed. After 200 epochs, the validation accuracy reaches 95.5$\%$. The training time is roughly 2 hours on a Tesla P100 GPU (16GB).

\subsection{Classification}
\begin{figure}[!htb]
    \centering
    \begin{tabular}{c}
        \includegraphics[scale=0.42, trim={0cm 0cm 0cm 0cm}, clip]{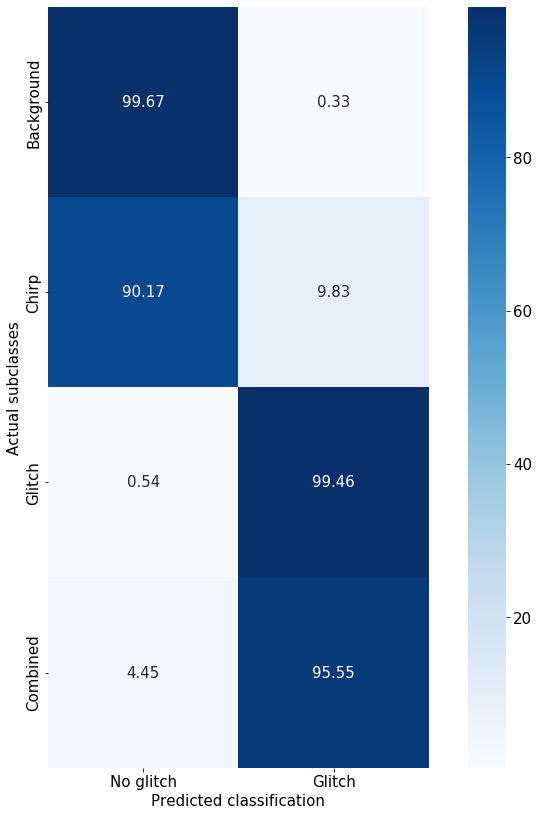}
    \end{tabular}
    \caption{Confusion matrix for the glitch label. The test has been conducted on 2812 TF maps showing glitches (744 glitch images and 2068 combined images) and 12000 that do not include any glitch (6000 background images and 6000 chirp images).}
    \label{confusion_matrix_glitch}
\end{figure}

To assess the identification of glitches, we run the classifier on the remaining images of each class, namely 6000 for the background and chirp class, 744 for the glitch class and 2068 for the combined class. The confusion matrix is shown in Figure \ref{confusion_matrix_glitch}. The threshold value to decide whether a TF map contain a glitch is chosen to be 0.5. Glitches appearing in the data are found with an accuracy above 95.55$\%$ while background and chirp images are correctly identified in at least 90.17 $\%$ of the cases. Note that the false-alarm rate for background images is very low, with roughly 0.33$\%$ of misclassified TF maps.

\section{Large-scale tests}\label{section:burst_search}
\subsection{Background analysis}

To test if the trained CNN can be used to reduce the false-alarm rate of ALBUS during a real search, we simulated a 5-year background search, accounting for 157772 time-frequency maps to process. The background is produced via time slides \cite{time_slides} with real data from Hanford and Livingston from the O3a run. Every image is passed to ALBUS to filter the background components. Its output is then used to evaluate the anomaly score of the map and finally passed through the glitch classifier. If the glitch probability is above 0.5, the TF map is classified as containing a glitch. \\

\begin{figure}[!htb]
    \centering
    \includegraphics[trim={0.1cm 0cm 0cm 0cm}, clip, scale=0.35]{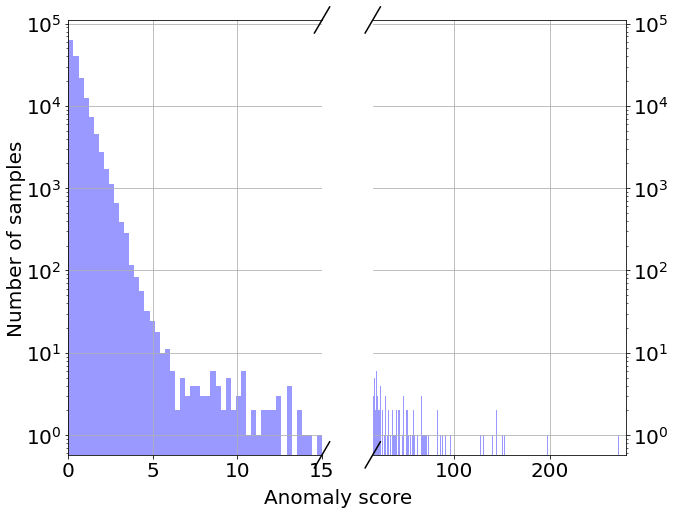}
    \caption{Histogram of the anomaly scores for 5 years of background.}
    \label{background_large}
\end{figure}

The background distribution of the anomaly scores is shown in Figure \ref{background_large}. As most of the images show a small anomaly score, these will not limit our sensitivity to burst signals. However, some background candidates get a score above 6 and should be examined. \\

\begin{figure}[!htb]
    \centering
    \includegraphics[trim={0.1cm 0cm 0cm 0cm}, clip, scale=0.38]{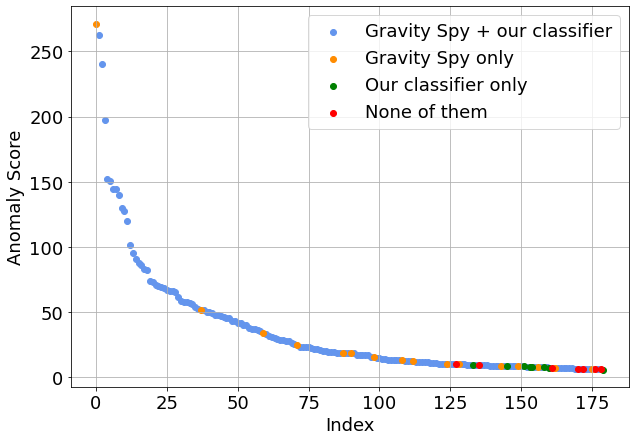}
    \caption{Classification of the 180 highest background candidates.}
    \label{background_highest}
\end{figure}

\begin{figure*}[htb]
 \centering
 \includegraphics[trim={0cm 0cm 0cm 0cm}, clip, scale=0.52]{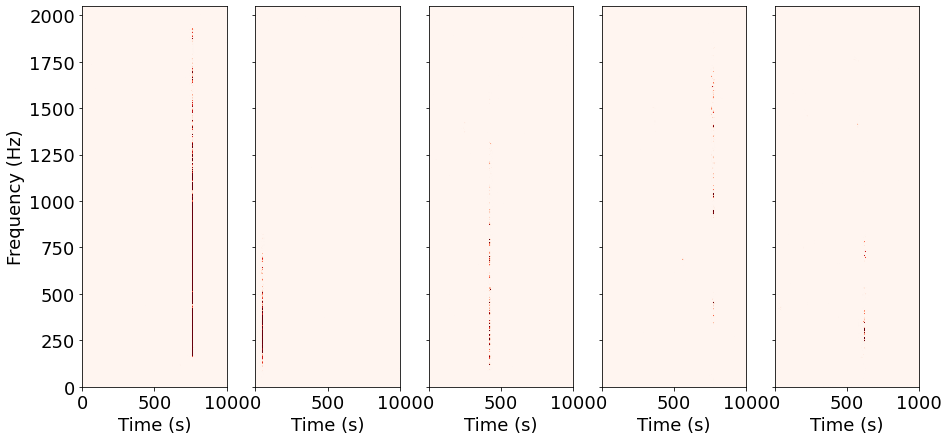}
 \caption{Outputs of ALBUS for TF maps in which Gravity Spy has identified a glitch while our classifier has not detected any. }
 \label{gravity_spy_only}
\end{figure*}

Among the highest candidates, we expect to find a majority of cross-correlated glitches. To compare the classifier with state-of-the-art glitch retrieval procedures, we use Gravity Spy \cite{GravitySpy}. For every image, we check if Gravity Spy has recorded a glitch either in the Hanford or Livingston data at that time. All the TF maps showing an anomaly score above 6 (180 in total) have been analyzed and their classification as images containing a glitch is shown in Figure \ref{background_highest}. Gravity Spy retrieves 165 glitches while our classifier identifies 157 of them, having 149 glitches in common. Gravity Spy cancels out candidates with a high anomaly score but some of them (16 in total) are missed by our CNN. The output of ALBUS for some of these TF maps is shown in Figure \ref{gravity_spy_only}. The glitches shown look like classical glitches although they present a higher minimal frequency compared to those appearing in Figures \ref{example_glitches} and \ref{example_chirp_glitch}. A probable explanation is that our classifier is sensitive to the bandwidth of the signals. By cross-correlating only 7 classes of glitches, we have limited the variability in the resulting TF map, somehow indirectly impacting the detection capability of the network. \\

\begin{figure*}[htb]
 \centering
 \includegraphics[trim={0cm 0cm 0cm 0cm}, clip, scale=0.52]{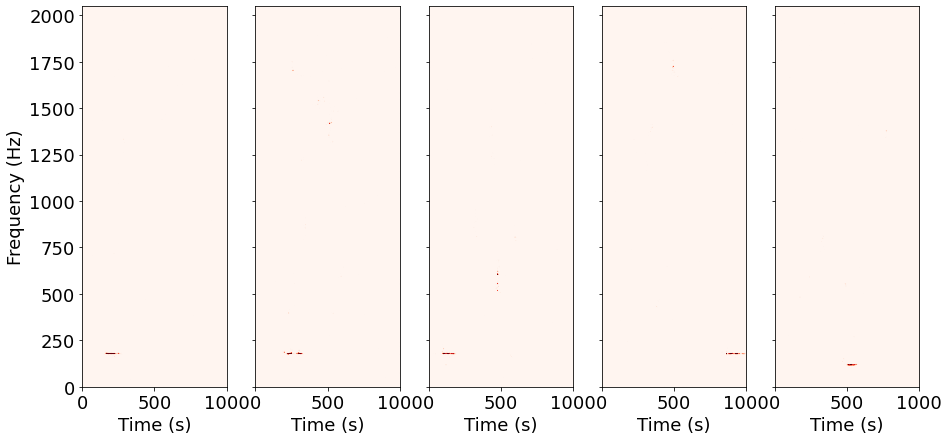}
 \caption{Outputs of ALBUS for TF maps in which our classifier has identified a glitch while Gravity Spy has not detected any.}
 \label{classifier_only}
\end{figure*}

\begin{figure}[!htb]
    \centering
    \includegraphics[scale=0.27, trim={0.1cm 0cm 0cm 0cm}, clip]{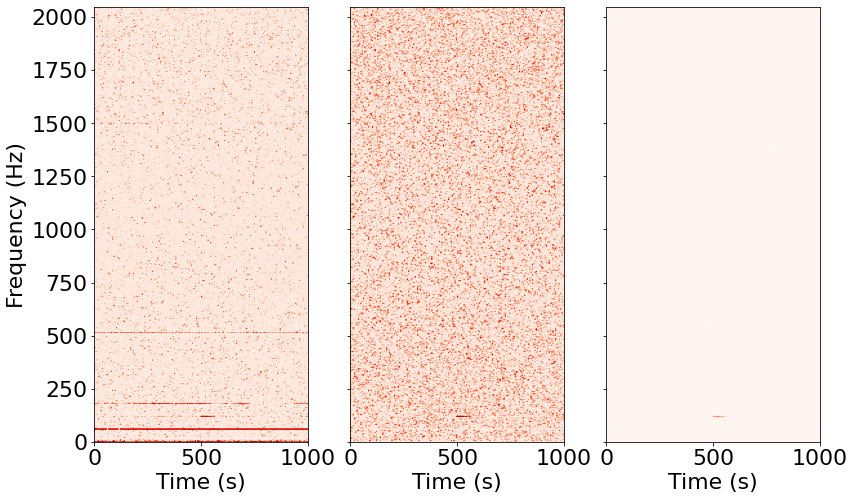}
    \caption{Example of artefact detected by our classifier. The left and central panel shows respectively the generated spectrogram before and after the whitening procedure while the right panel shows the output of ALBUS.}
    \label{check_albus}
\end{figure}

On the other hand, the classifier recognizes 8 glitches at low anomaly scores (green dots in Figure \ref{background_highest}) for which Gravity Spy does not detect anything. Since Gravity Spy takes Omicron \cite{Omicron} triggers as input, the latter might not have produced triggers for these particular 8 events. Figure \ref{classifier_only} shows the output of ALBUS, i.e. the input of the classifier, for 5 of them. The patterns are narrowband and last 100 to 150 seconds. As these artefacts appear as mono-frequency lines, they might be related to the power line at 60 Hz in the United-States and its harmonics \cite{detchar2021}. Figure \ref{check_albus} shows the spectrogram before and after the whitening for one of the example in Figure \ref{classifier_only} and reveals that the first harmonic (120 Hz) barely appears in the data. Therefore the whitening procedure could not clean that power line. 

\subsection{Burst signal analysis}
To complete the tests carried out on background images, we applied our CNN to 4 expected types of long duration burst signals across 22 Hrss intensities. In order to claim a detection, the output of ALBUS should contain an anomaly but it cannot be a glitch. The second condition is met when the glitch probability (GP) is lower than 0.5. To validate the first condition, the anomaly score (AS) of the output map should be higher than the anomaly score obtained for background images. In Figure \ref{background_highest}, the highest anomaly score that is not identified as a glitch by Gravity Spy for a background map is 10.24. The two thresholds used for the analysis are therefore:
\begin{equation}
    GP < 0.5
\end{equation}
\begin{equation}
    AS > 10.24
\end{equation}

\begin{figure*}[!htb]
    \centering
    \includegraphics[scale=0.6, trim={0.1cm 0cm 0cm 0cm}, clip]{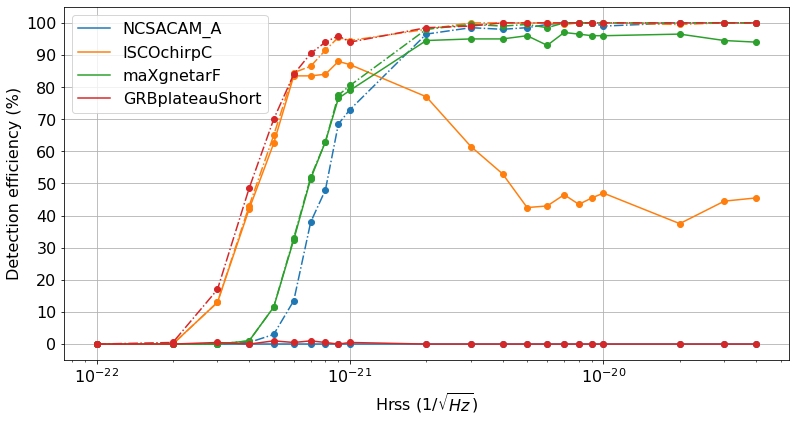}
    \caption{Detection efficiency for long duration waveforms (GRBplateauShort \cite{GRBplateauShort}, ISCOchirpC \cite{ISCOchirp}, maXgnetarF \cite{magnetar_waveform}, NCSACAM-A \cite{ecbc}) The dash-dotted curves refer to the scenario where the only threshold is the anomaly score while the glitch probability is also used in the case of the continuous lines.}
    \label{detection_efficiency}
\end{figure*}

\noindent
Figure \ref{detection_efficiency} shows the efficiency curves for 4 different waveforms in two different scenarios. Every dot is the estimation over 200 injections performed at the same hrss intensity. \\

\begin{figure*}[!htb]
    \centering
    \begin{tabular}{cc}
        \includegraphics[scale=0.32, trim={0.1cm 0cm 0cm 0cm},clip]{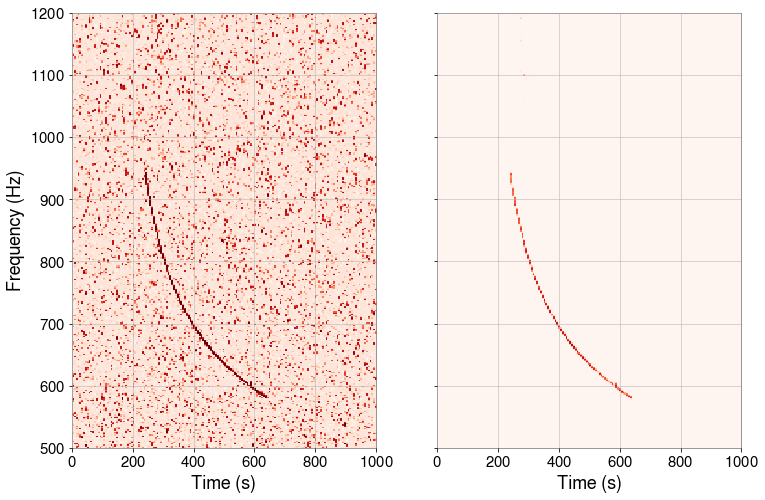} &
        \includegraphics[scale=0.32, trim={0.1cm 0cm 0cm 0cm},clip]{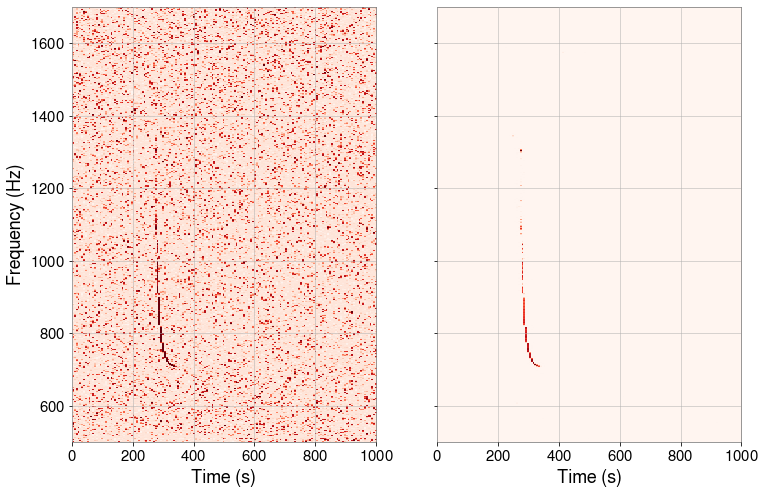}
    \end{tabular}
    \begin{tabular}{cc}
        \includegraphics[scale=0.32, trim={0.1cm 0cm 0cm 0cm},clip]{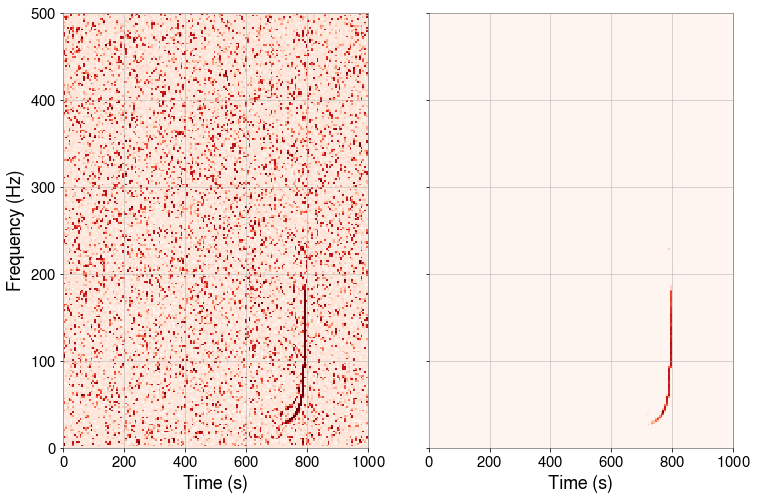} &
        \includegraphics[scale=0.32, trim={0.1cm 0cm 0cm 0cm},clip]{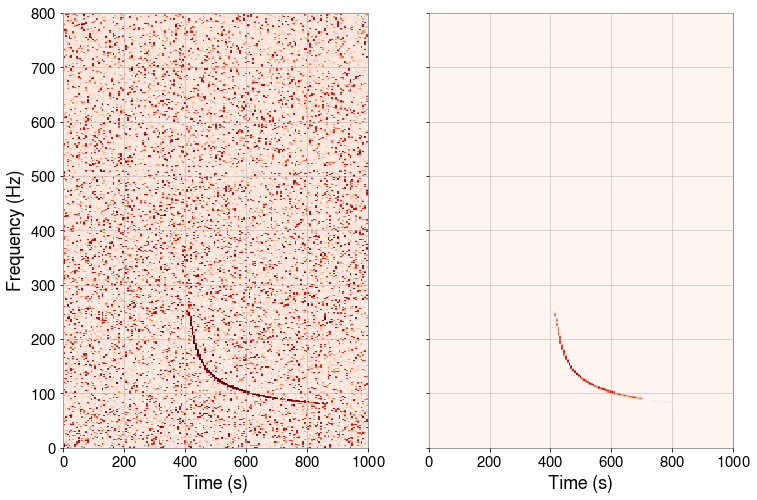}
    \end{tabular}
    \caption{Examples of detection performance on long-duration waveforms (top left: \textit{maXgnetar-D} \cite{magnetar_waveform}, top right: \textit{ISCOchirp-C} \cite{ISCOchirp}, bottom left: \textit{NCSACAM-A} \cite{ecbc} and bottom right: \textit{GRBplateau} \cite{GRBplateauShort}). The left image of each panel is the input TF map and the right panel shows the output of ALBUS.}
    \label{examples detection}
\end{figure*}

The detection efficiency for the long duration waveforms is highly dependent on the shape of the footprint left in the TF maps. Figure \ref{examples detection} shows the pattern left by the 4 selected waveforms. The detection efficiency for the magnetar model is very similar whether if the glitch probability is used or not, meaning that the classifier does not recognize it as a glitch. This is not the case for the 3 other models. Indeed, our classifier identifies events of that kind as glitches most of the time, which is certainly due to their steep behavior. Moreover, these sorts of chirps might not be abundant enough in the data since the chirp generation parameters have been randomized. The classifier might therefore consider only the steep part to classify them as glitches. \\


\section{Discussion and Conclusion}\label{section:conclusion}

The anomaly score has been defined as a statistics to detect and rank signals in the output map of ALBUS. It can also be used as the unique detection threshold of our pipeline, showing encouraging results. Gravity Spy \cite{GravitySpy} could then be used to remove the false-alarms due to glitches. \\

In this paper, we have applied a convolutional neural network to the identification of detector glitches in the time-frequency space of the cross-correlated LIGO noise. The training has been carried out both with glitches and chirping signals to help the network learn their distinct morphologies. The network recognizes more than 95$\%$ of the glitches while it has a low false-alarm rate on random chirping signals. The performance of the classifier can be improved by adding more glitch classes to the training data, increasing accordingly the variability in their cross-correlation output. Indeed, we only select 7 glitch classes in this work, limiting the bandwidth diversity in the data. \\

In the same way, chirp data have to be adapted to improve the performances on long duration models showing a steep pattern in the TF space. The efficiency of our network can be improved by either overpopulating rapidly chirping signals in the data or by training directly on a subset of long duration waveforms. Future works will contribute to the improvement of the classifier introduced in this paper. \\

\section*{Acknowledgements}
The author thanks Grégory Baltus for useful discussions and comments. V.B. is supported by the Gravitational Wave Science (GWAS) grant funded by the French Community of Belgium. This material is based upon work supported by NSF's LIGO Laboratory which is a major facility fully funded by the National Science Foundation. The authors are grateful for computational resources provided by the LIGO Laboratory and supported by the National Science Foundation Grants No. PHY-0757058 and No. PHY-0823459. \\

\section*{Appendix}
\subsection{Visibility levels}\label{appendix1}

\begin{figure*}[hbt!]
    \centering
    \begin{tabular}{ccc}
        \includegraphics[scale=0.4]{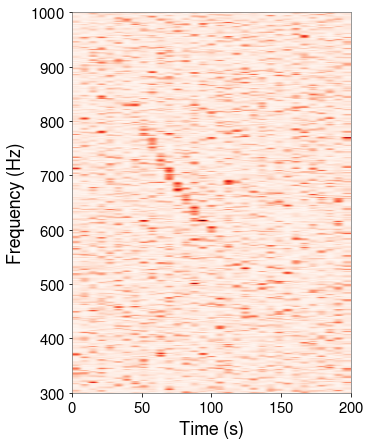} &
        \includegraphics[scale=0.4]{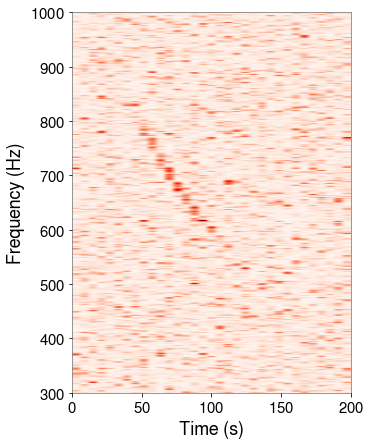} &
        \includegraphics[scale=0.4]{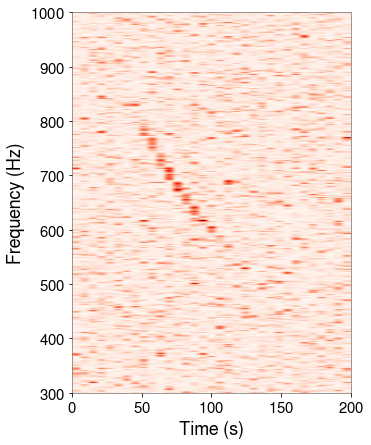}
    \end{tabular}
    \begin{tabular}{ccc}
        \includegraphics[scale=0.4]{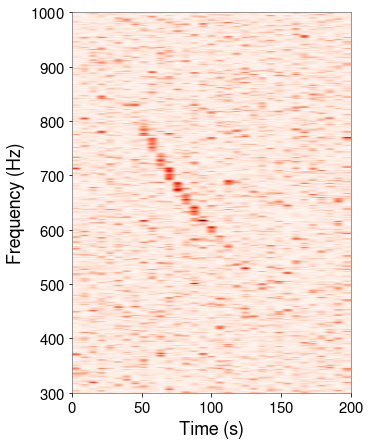} &
        \includegraphics[scale=0.4]{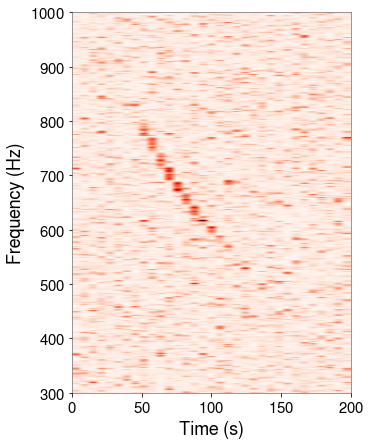} &
        \includegraphics[scale=0.4]{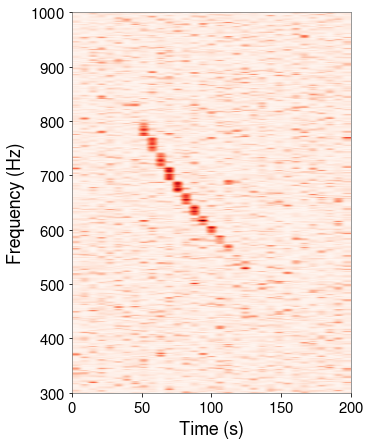}
    \end{tabular}
    \begin{tabular}{ccc}
        \includegraphics[scale=0.4]{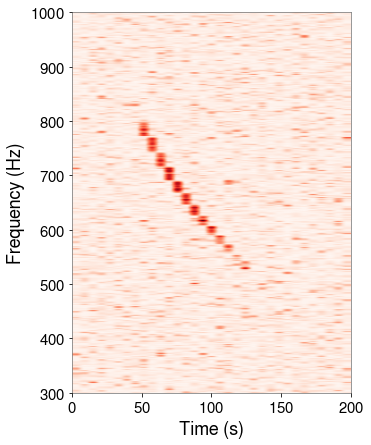} &
        \includegraphics[scale=0.4]{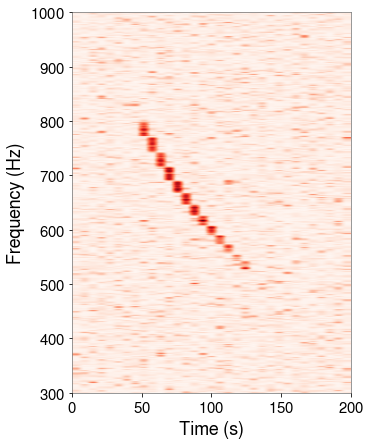} &
        \includegraphics[scale=0.4]{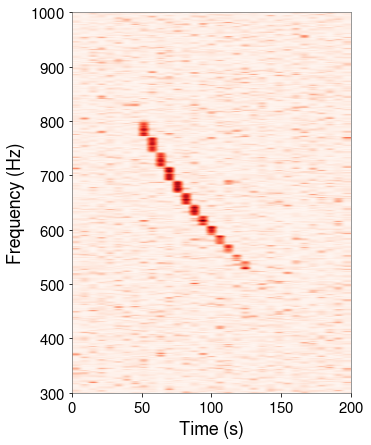}
    \end{tabular}
    \caption{Visibility levels used in this work shown through a unique injected chirp. The values are, from top left to bottom right, 12, 14, 16, 18, 20, 30, 40, 50 and 60.}
    \label{visibility_levels}
\end{figure*}

Figure \ref{visibility_levels} shows the 9 levels of visibility that have been used to inject chirp signals in the chirp and combined training sets.

\bibliographystyle{apsrev}
\bibliography{main.bib}

\end{document}